\documentclass{aa}  

\usepackage{graphicx}
\usepackage{txfonts}
\usepackage{color}

\begin{document}

   \title{A survey of HI gas toward the Andromeda Galaxy}

 \author{J. Kerp
          \inst{1}
          \and
          P.M.W. Kalberla\inst{1}
          \and
          N. Ben Bekhti\inst{1}
\and
          L. Fl\"oer\inst{1}
\and
        D. Lenz\inst{1}
\and
        B. Winkel\inst{2,1}
          }

   \institute{Argelander-Institut f\"ur Astronomie
              Auf dem H\"ugel 71, D-53121 Bonn, Germany\\
              \email{jkerp@astro.uni-bonn.de}
         \and
             Max-Planck-Institut f\"ur Radioastronomie, Auf dem H\"ugel 69, D-53121 Bonn, Germany\\
             }

   \date{Received September 15, 1996; accepted March 16, 1997}

 
  \abstract
   {The subsequent coalescence of low--mass halos over cosmic time is thought to be the major formation 
channel of massive spiral galaxies like the Milky Way and the Andromeda Galaxy (M31). The gaseous halo of a massive galaxy is considered to be the reservoir of baryonic matter persistently fueling the star formation in the disk.
Because of its proximity, M31 is the ideal object for studying the structure of the halo gas in great detail.
   }
   {Using the latest neutral atomic hydrogen (HI) data of the Effelsberg-Bonn HI Survey (EBHIS) allows comprising a comprehensive inventory of gas associated with M31. The primary aim is to differentiate between physical structures belonging to the Milky Way Galaxy and M31 and accordingly to test the presence of a M31 neutral gaseous halo.
   }
   {Analyzing the spatially fully sampled EBHIS data makes it feasible to trace coherent HI structures in space and radial velocity. To disentangle Milky Way and M31 HI emission we use a new approach, along with the traditional path of setting
an upper radial velocity limit, by calculating a difference second moment map.
   }
   {We argue that M31's disk is physically connected to an asymmetric HI halo of tens of kpc size, the M31\,cloud. We confirm the presence of a coherent low--velocity HI filament located in between M31 and M33 aligned at the sky with the clouds at systemic velocity. The physical parameters of the HI filament are comparable to those of the HI clouds at systemic velocity. We also detected an irregularly shaped HI cloud that is is positionally located close to but offset from the stellar body of And\,XIX.
   }
{}


   \maketitle
%

\section{Introduction}

The Local Group (LG) of galaxies is a unique laboratory for testing and investigating the formation and evolution of stellar and gaseous bodies. Already today, HI data of LG objects offer a sufficiently high-mass sensitivity to detect gaseous low mass halos of about $M_{\rm HI} \sim 10^4\,{\rm M_\odot}$ \citep{Thilker2004, Westmeier2008}, which are orders of magnitude below those of the ``building blocks'' of massive baryonic halos, the dwarf galaxies \citep{Blitz1999,Klypin1999}.
According to standard $\Lambda$CDM cosmology, we expect to detect massive concentrations of baryons (large spiral
galaxies) in the
current epoch. They are closely correlated spatially with a high number of these low
mass halos \citep{Klypin1999}.

Because of the complex spatial and kinematical mixture of the individual gas clouds in the Milky Way galaxy and its halo, these building blocks are difficult to identify. Considering the strong environmental dependence of the chemical composition, as well as the heating and cooling of the different gaseous phases \citep{Wolfire2003}, the quantitative modeling demands a very careful approach \citep{KalberlaKerp2009}. Consequently, diffusely distributed low column density HI structures located in the Milky Way halo and in the disk-halo interface are difficult to separate from the luminous and highly dynamic disk gas emission if not separated well in Doppler velocity.

Because of their proximity and partly significant separation in radial velocity from the Milky Way HI emission, the Andromeda (Messier 31, M31) and Triangulum (Messier 33, M33) galaxies allow investigation of diffuse low mass gaseous halo structures. Here, we study the 21-cm line emission of neutral atomic hydrogen (HI) observed within the course of the Effelsberg Bonn HI Survey \citep[][EBHIS]{Winkel2015, Kerp2011, Winkel2010}.

\citet{Blitz1999} performed a comprehensive observational inventory of the LG galaxies and the high--velocity clouds (HVCs). HVCs are proposed to populate the circumgalactic environment of the massive galaxies \citep{Westmeier2008}. Their radial velocities deviate from the systemic velocities of these galaxies up to a few hundred ${\rm km\,s^{-1}}$.
In the case of M31, all investigations are faced with confusion at lower radial velocities than the systemic ones \citep[see][Fig.\,2]{Lehner2015} between M31 and Milky Way HI.
Distinguishing between M\,31 and the unrelated Milky Way foreground, HI emission, in particular with HVC emission \citep{Simon2006}, is a major issue that we consider here.

To minimize systematic biases due to this Milky Way gas confusion, previous HI studies of M31 and its environment either focused on structures at M31's systemic velocities
\citep{BraunThilker2004,Wolfe2013} around $v_{\rm LSR} = -300\,{\rm
  km\,s^{-1}}$ or instead used upper radial velocity thresholds
\citep{Thilker2004,Westmeier2005} ranging between $-160\,{\rm
  km\,s^{-1}}\,\leq\,v_{\rm LSR} \leq -140\,{\rm km\,s^{-1}}$. For the
northern portion of the Leiden/Argentine/Bonn survey data
\citep[][LAB]{LABsurvey}, the Leiden--Dwingeloo survey
\citep[][LDS]{hartmannburton}, \citet{Blitz1999} extended the analyzed radial
velocity range up to $ v_{\rm LSR} \leq -85\,{\rm km\,s^{-1}}$ and
identified the ``M31\,cloud'' apparently connected in HI emission to M31's disk.

Not only their finding but also the studies of
M31's dwarf satellite system \citep{Collins2013} motivate us to inspect this radial velocity regime more closely. 
For example, the dwarf spheroidal satellite galaxies And X
\citep{Kalirai2010}, XXIV, XXV, and XIX \citep{Collins2013} populate the
radial velocity range $-164\,{\rm km\,s^{-1}}\,\leq\,v_{\rm
  LSR}\,\leq\,-107\,{\rm km\,s^{-1}}$, which has so far only been
marginally explored by HI observations. 
The lack of spatial correlation between sensitive HI observations
\citep{Thilker2004,Braun2009} and the Pan-Andromeda Archeological Survey
(PAndAS, \citet{McConnachie2009}) optical data ``on all scales''
\citep{Lewis2013} might be considered as a result of the applied
HI radial velocity limits. 

We start our analysis by exploring the EBHIS data following the classical path of setting an upper radial velocity limit of $v_{\rm LSR} = -95\,{\rm km\,s^{-1}}$. Furthermore, we invent a new approach by calculating a so--called ``difference second moment map'' to disentangle coherent HI structures in the space and the velocity domain.
Essential for our investigation is the sufficiently high spatial resolution that is offered by EBHIS \citep{Kerp2011, Winkel2015}. Using the difference second moment map approach, we analyze M\,31's HI emission up to $v_{\rm LSR} = -25\,{\rm km\,s^{-1}}$.

The paper is organized in six sections. Section 2 describes the EBHIS data briefly. Section 3 compiles our results of a morphological investigation and discusses the detection limits. It also comprises a comparison of the EBHIS data with previous single-dish HI observations of M31. Section 4 presents our new approach of calculating difference second moment maps to distinguish between M\,31 and unrelated Milky Way HI emission. Section 5 contains a discussion of the physical association of the reported gaseous structures with the Andromeda galaxy and its dwarf galaxy system. Here, the impact of the tidal interaction, galaxy impact, and proper motion of M31 as an entity against a hypothetical LG medium is discussed. Section 6 presents our summary and conclusions. 

\section{Data}
We use EBHIS 21-cm line data from the first coverage \citep{Kerp2011, Winkel2015, Winkel2010}. EBHIS covers the whole northern sky
above a declination of $-5^\circ$ with a radial velocity range
$-1000\,\leq\,v_{\rm LSR}{\rm [km\,s^{-1}]}\,\leq\,19000$.  
The EBHIS data of M31 have been extracted and processed by the standard data reduction pipeline \citep{Winkel2010, Winkel2015}. The median root-mean-square (rms) uncertainties of the antenna brightness temperature is 90\,mK for a spectral channel separation of $\Delta v_{\rm LSR} = 1.29\,{\rm km\,s^{-1}}$. To increase the signal--to--noise
ratio for our M31 analysis, we binned spectral channels by a factor of
ten. The contamination by random noise was minimized in the map presentation by generating a
mask with a $5\sigma$ threshold, equivalent to 142\,mK across the whole
field of interest. The EBHIS angular resolution of 10.8\,arcmin
(2.36\,kpc at the M31 distance of 752\,kpc \citep{Riess2012}) is
unaffected by this filtering process.

   \begin{figure*}
     \centerline{
        \includegraphics[scale=0.18, angle=0]{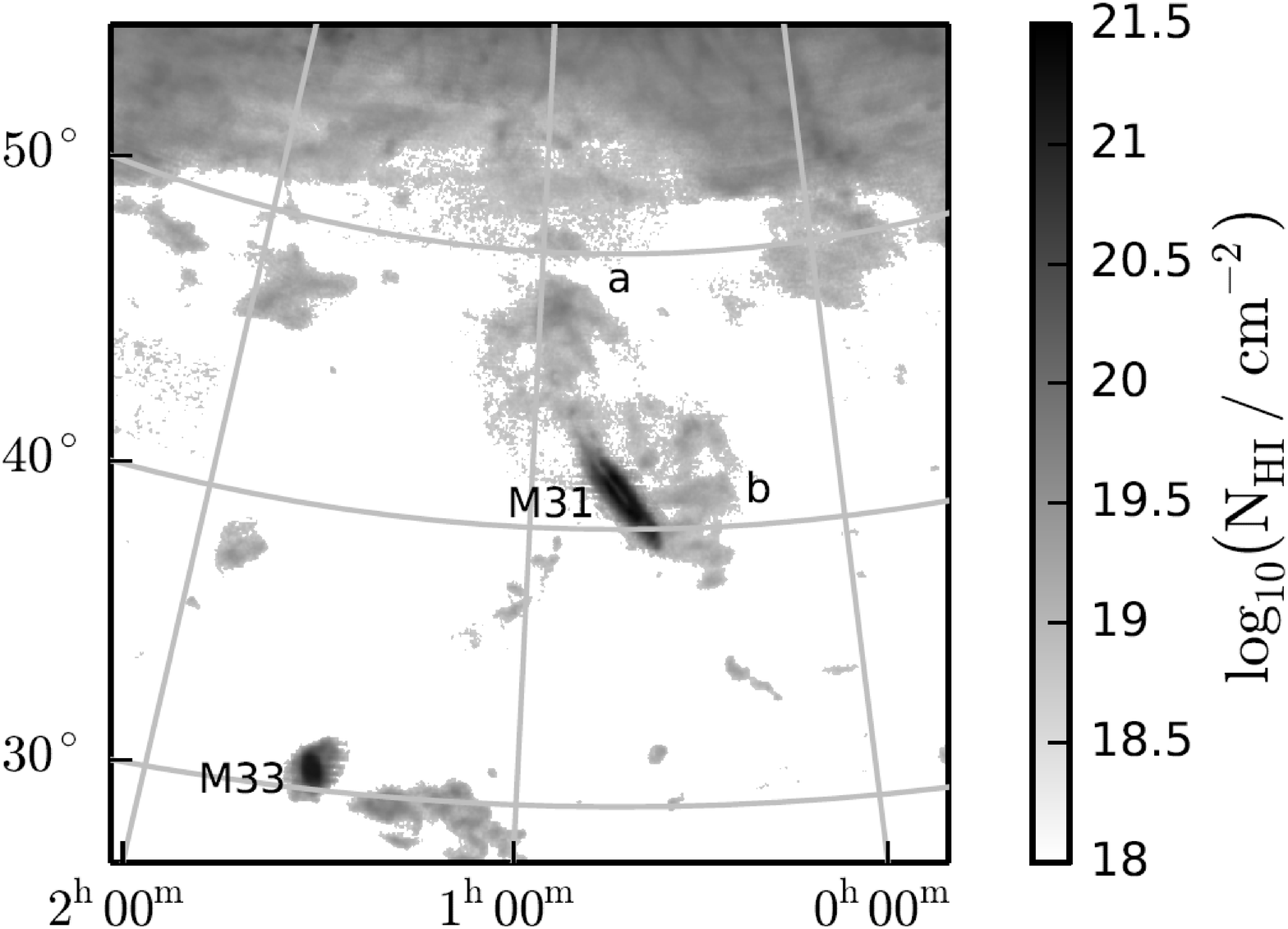}
        \includegraphics[scale=0.18, angle=0]{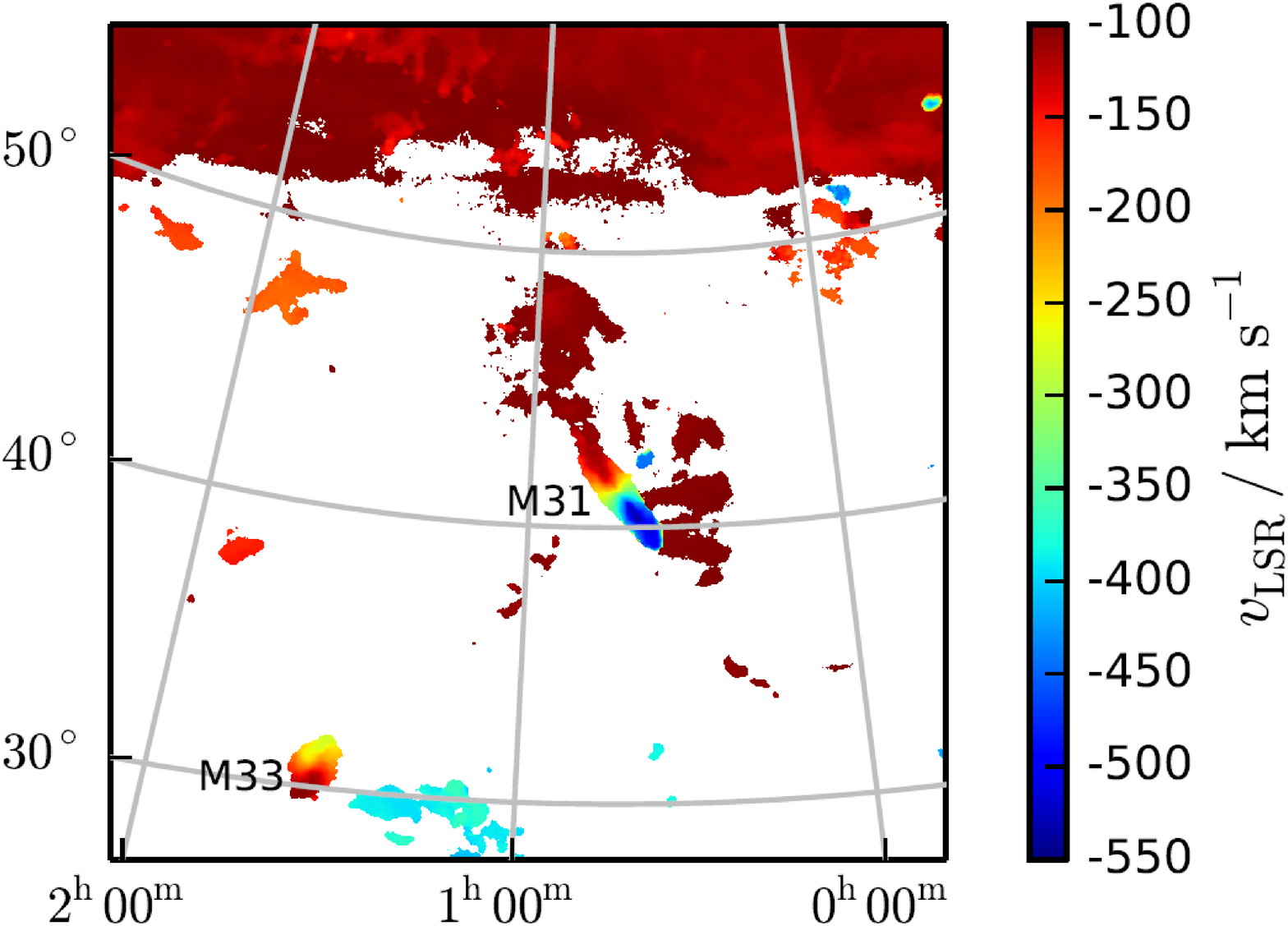}
                }
      \caption{{Integrated column density and Doppler velocity map of M\,31 and its environment.} {\it Left panel:} HI emission exceeding the EBHIS 5-$\sigma$ sensitivity threshold across the radial velocity range $-600\,{\rm km\,s^{-1}}\,\leq\,v_{\rm LSR} \,\leq\,-95\,{\rm km\,s^{-1}}$ is displayed.
The plane of the Milky Way galaxy is visible along the top, and M33 is seen in the lower left corner. At the distance of M31, the extent of the map corresponds to 400\,kpc.
{\it Right panel:} the intensity weighted Doppler velocity map of the area of interest. Using the radial velocity information, it is feasible to distinguish between individual coherent HI structures. The rotation of both local group spiral galaxies, as well as HI emission of so-called high-velocity clouds complex H \citep{Simon2006, Blitz1999} ($v_{\rm LSR} \simeq -200\,{\rm km\,s^{-1}}$) next to the Milky Way galaxy emission, can be identified.}
    \label{Fig:mom0}

    \end{figure*}

\section{Results and comparison with previous studies}

Figure\,\ref{Fig:mom0} (left panel) displays a column density map. M\,31 is located close to the center of the map. The HI emission from the Milky Way plane is visible along the top, while M\,33 is located in the lower left corner.
 The extent of the map is about $30^\circ$, corresponding to $\sim$\,400\,kpc at the M31 distance of 752\,kpc \citep{Riess2012}.
The right panel of Figure\,\ref{Fig:mom0} shows the corresponding radial velocity map of the area of interest (see Sect.\ref{Sect:Separation}, Eq.\,\ref{Eq:mom1}). Coded in colors are the intensity-weighted Doppler velocities of the gaseous structures. These first moment maps allow us to distinguish kinematically between individual coherent HI structures.

 \begin{figure*}
     \centerline{
        \includegraphics[scale=0.5, angle=0]{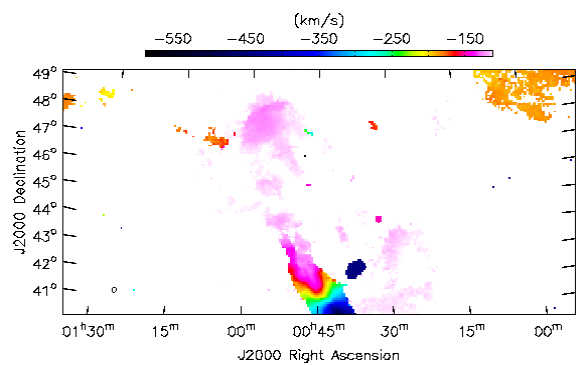}
                        }
      \caption{First moment map toward the northern portion of the M31 HI gas distribution.  Displayed are radial velocities in the range $-588\,\leq\,v_{\rm LSR}\,{\rm [km\,s^{-1}]}\,\leq\,-110$. Here, the confusion of M31 HI gas with that of HVC complex H is most severe. This first moment map allows HI gas superimposed on a single line of sight but moving at different radial velocities to
be disentangled.}
    \label{Fig:mom1_detail}

\end{figure*}

To illustrate this in more detail, we investigated the superposition of the HI gas emission toward the northern portion of M31, in particular with HVC complex H \citep{Simon2006}.
Figure\,\ref{Fig:mom1_detail} displays the column density weighted velocity structure toward this particular area of interest. The HI gas appears to be subdivided in individual HI structures ranging in angular size between a few tens of arcmin to several degrees. All these individual HI objects are showing up as coherent HI structures in radial velocity. Figure\,\ref{Fig:mom1_detail} displays a continuous connection in column density and radial velocity between the M31 disk and the large extended HI structure labeled with (a) in Fig.\ref{Fig:mom0}.

In the following we focus on those diffuse HI emission structures that are prominent at Doppler velocities of $v_{\rm LSR} \simeq -115\,{\rm km\,s^{-1}}$ (labeled in Fig.\,\ref{Fig:mom0}, left panel with (a) and (b)). They appear to emerge from M31's HI disk.
We also identified at these low radial velocities HI components positionally located close to the M31/M33 HI bridge at systemic velocities \citep{BraunThilker2004,Wolfe2013}.

\subsection{Leiden--Dwingeloo Survey: ``M31\,cloud''}
A comparison of our EBHIS with low--angular resolution LDS data by \citet[see Fig.\,5]{Blitz1999} shows very good agreement.
\citet{Blitz1999} labeled the whole HI structure positionally offset from M\,31's disk with a total extent of about 14 degrees as the ``M31\,cloud''.  They discuss the striking
morphology and positional agreement of the M31\,cloud with M31 but felt
unable to derive clear evidence of any association of the M31\,cloud with M31.  

\subsection{Green Bank and Effelsberg Telescopes: M31-HVCs} 
In a dedicated sensitive HI survey with the Green-Bank Telescope (GBT), \citet{Thilker2004} mapped a $7^\circ \times 7^\circ$ region centered on the stellar body of M31. Their Fig.\,2 displays the HI column density distribution, which they attribute to the M31-HVC
gas. \citet{Thilker2004} calculated a spatial mask for the brightness temperature using a velocity threshold of $v_{\rm LSR} \leq -170\,{\rm km\,s^{-1}}$ to distinguish between M31 HI disk and unrelated Milky Way emission. As the consequence of the applied method, the M31 halo gas appears to be disconnected from M\,31's HI disk.
Because our Fig.\,\ref{Fig:mom0}  displays the full observed HI emission up to $v_{\rm LSR} = -95\,{\rm km\,s^{-1}}$, we obtain a markedly different presentation. In our maps, it is possible to identify smooth, continuously connected HI filaments emerging from M31's disk instead of isolated HVCs as described by \citet{Thilker2004}. 

\subsection{Distribution of M\,31--HVCs}
\label{Lopsect}

Figure 2 of \citet{Thilker2004}
shows M\,31-HVCs with comparable HI column densities above and below the M\,31's disk.
The HI distribution above and below M31's stellar disk in our EBHIS maps (Fig.\,\ref{Fig:mom0}) but also in the LDS map investigated by \citet{Blitz1999} appears to be very different from these GBT results. Both display smooth and coherent HI structures rather than individual filaments.
Next, we investigate whether the different selection criteria or sensitivity limitations of the all-sky survey data map cause this apparent asymmetry. For this aim we study the southeastern disk side in more detail.

\begin{figure*}
     \centerline{
        \includegraphics[scale=0.5, angle=0]{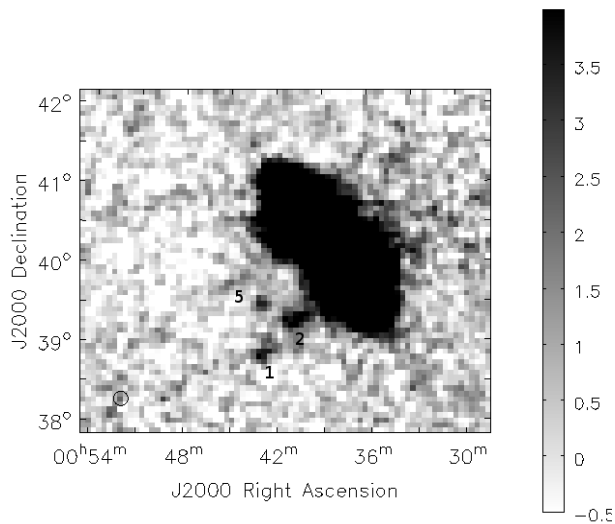}
                        }
      \caption{Integrated ($-530\, \leq\, v_{\rm LSR}\,{\rm [ km\,s^{-1}]}\,\leq-482$) brightness temperature $[{\rm K\,km\,s^{-1}}]$ map of the area toward Andromeda's southeastern disk. The M31-HVCs 1,2 and 5 discovered by \citet{Westmeier2008} in that radial velocity range are also detected by EBHIS, however the EBHIS column densities are lower by about 25\%. These HVCs are below the applied $5-\sigma$ threshold applied to the other maps. To demonstrate the baseline quality of the EBHIS standard data reduction the lower brightness temperature cut has been set to $-0.5$\,${\rm K\,km\,s^{-1}}$.}
    \label{Fig:M31HVCs}

\end{figure*}

The Effelsberg maps presented by \citet{Westmeier2008} show M31-HVCs located below the southeastern portion of M31's disk. In Fig.\,\ref{Fig:mom0} this portion of the M31 disk appears to be featureless. According to \citet{Westmeier2008}, M31-HVCs HI column densities are found to be about $N_{\rm HI} = 1\cdot 10^{19}\,{\rm cm^{-2}}$.
Figure\,\ref{Fig:M31HVCs} displays the EBHIS HI column density map of
M31's gas distribution integrated across the relevant radial velocity range
$-530\,\leq\,v_{\rm LSR}\,[{\rm km\,s^{-1}}]\,\leq-480$. The M31-HVCs 1, 2, and 5, according to the naming convention of \citet{Westmeier2008}, are significantly detected by EBHIS. The EBHIS column densities are about 25\% lower than those reported by \citet{Westmeier2008}. We attribute this difference to residual baseline uncertainties of the EBHIS data.
However, the detection of these faint M31-HVCs in the shallow EBHIS data demonstrates the high quality of the EBHIS data processing pipeline \citep{Winkel2015, Winkel2010, Martin2015}.

The different appearance between the northern and
southern portions of M31 HI disk in the EBHIS data is significant. It is not the result of artifacts of the data processing.
The EBHIS data is consistent with an extended and lopsided HI structure located toward the northwestern portion of the M31 disk identified first by \citet{Blitz1999}.

\subsection{Testing the association between M31 and the M31\,cloud}
\begin{figure*}
     \centerline{
        \includegraphics[scale=0.8, angle=0]{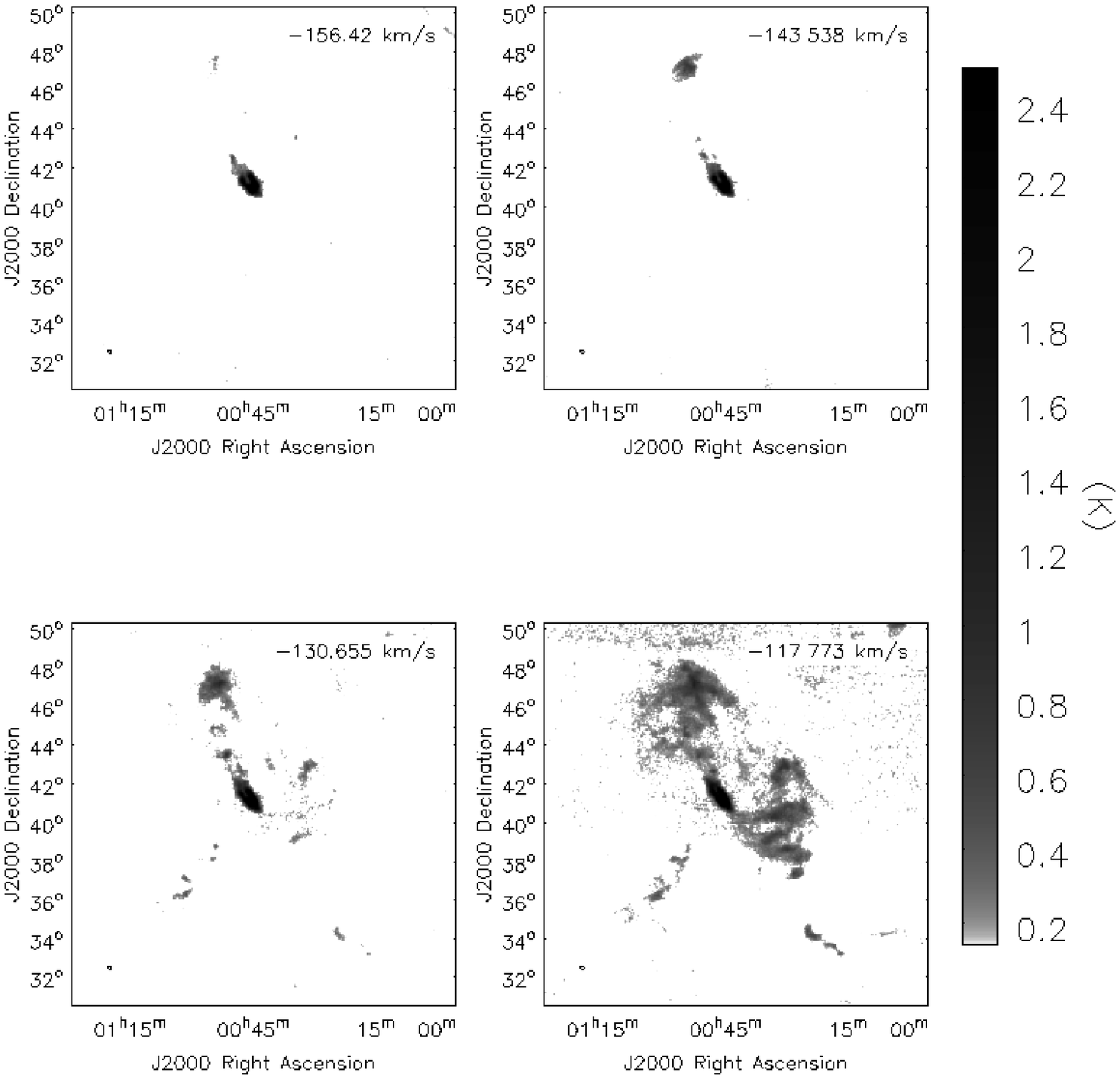}
                        }
      \caption{Brightness temperature maps of the area of interest for four subsequent velocity channels ($\Delta v_{\rm LSR} = 12.9\,{\rm km\,s^{-1}}$). Displayed is the HI emission exceeding the applied 5-$\sigma$ threshold of the EBHIS data. The insets in the top right corner of each panel gives the Doppler velocity relative to the local standard of rest. For display purposes, the brightness temperature is clipped above $T_{\rm B} = 2.5$\,K.}
    \label{Fig:fourpanels}

\end{figure*}

Figure\,\ref{Fig:fourpanels} displays the EBHIS brightness temperature distribution of four subsequent velocity channels. The top left panel shows the most negative velocity channel. Faint HI emission above the 5-$\sigma$ level is apparent, positionally consistent with the northern structure (a) in Fig.\,\ref{Fig:mom0}. The next channel presented in the top righthand panel reveals a much more extended HI emission patch. The bottom lefthand panel shows very extended HI emission in the northwestern halo portion of M31's disk and the HI peaks of the filamentary structure in between M31 and M33. The bottom righthand panel displays a continuous connection between M31's disk and these halo structures. This channel-by-channel ``evolution'' from a patchy to a coherent HI structure in the EBHIS data cube is suggestive of an association between M31 and the M31\,cloud. The brightest portions of these coherent HI structures are positionally coincident with the HI filaments observed with the GBT \citep{Thilker2004}. \citet{Thilker2004} attributed these filaments to individual isolated M31-HVCs.

\begin{figure*}
     \centerline{
        \includegraphics[scale=0.3, angle=0]{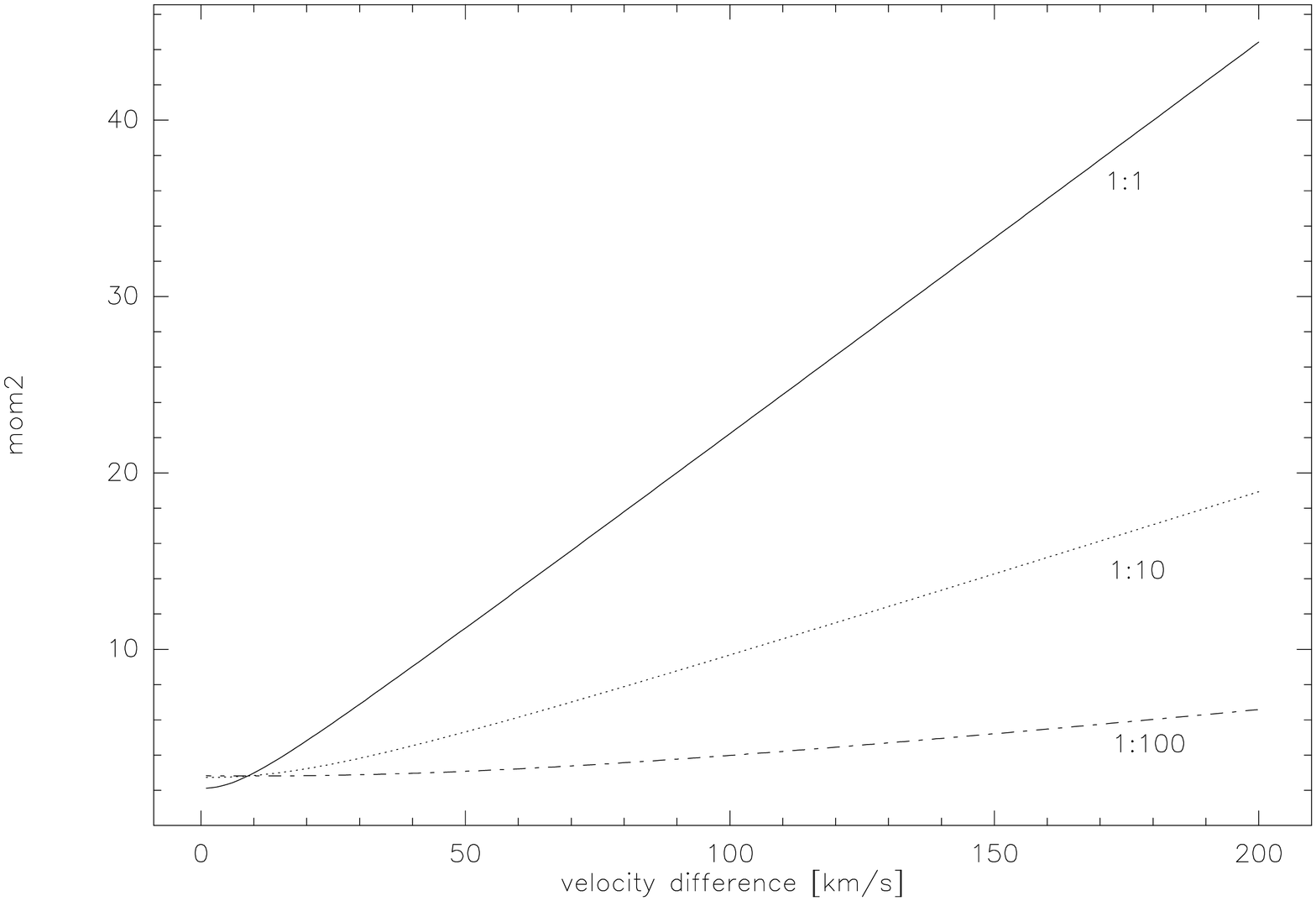}
                }
      \caption{Displayed are the second moments as a function of the radial velocity difference of two gas columns with the same velocity dispersion but different $N_{\rm HI}$. $1:1$ denotes the situation of a $N_{\rm HI}$ ratio of unity, $1:10$ and $1:100$ even higher column density ratios. The higher the contrast of both superimposed column densities, the larger the necessary separation in radial velocity for a significant separation of both clouds in $M_2$.}
    \label{Fig:coldencontrast}
\end{figure*}

\begin{figure*}
     \centerline{
              \includegraphics[scale=0.48, angle=0]{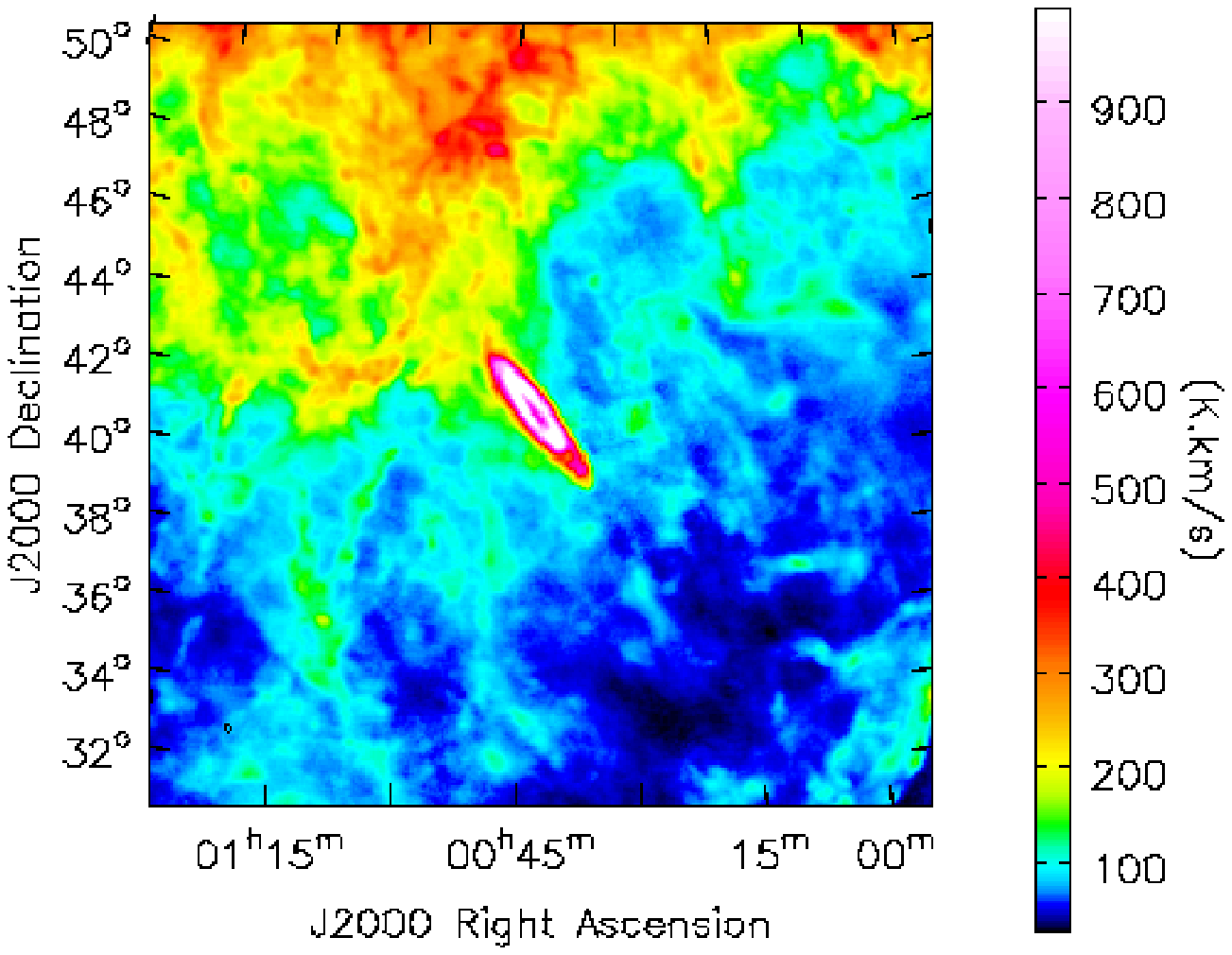}
        \includegraphics[scale=0.48, angle=0]{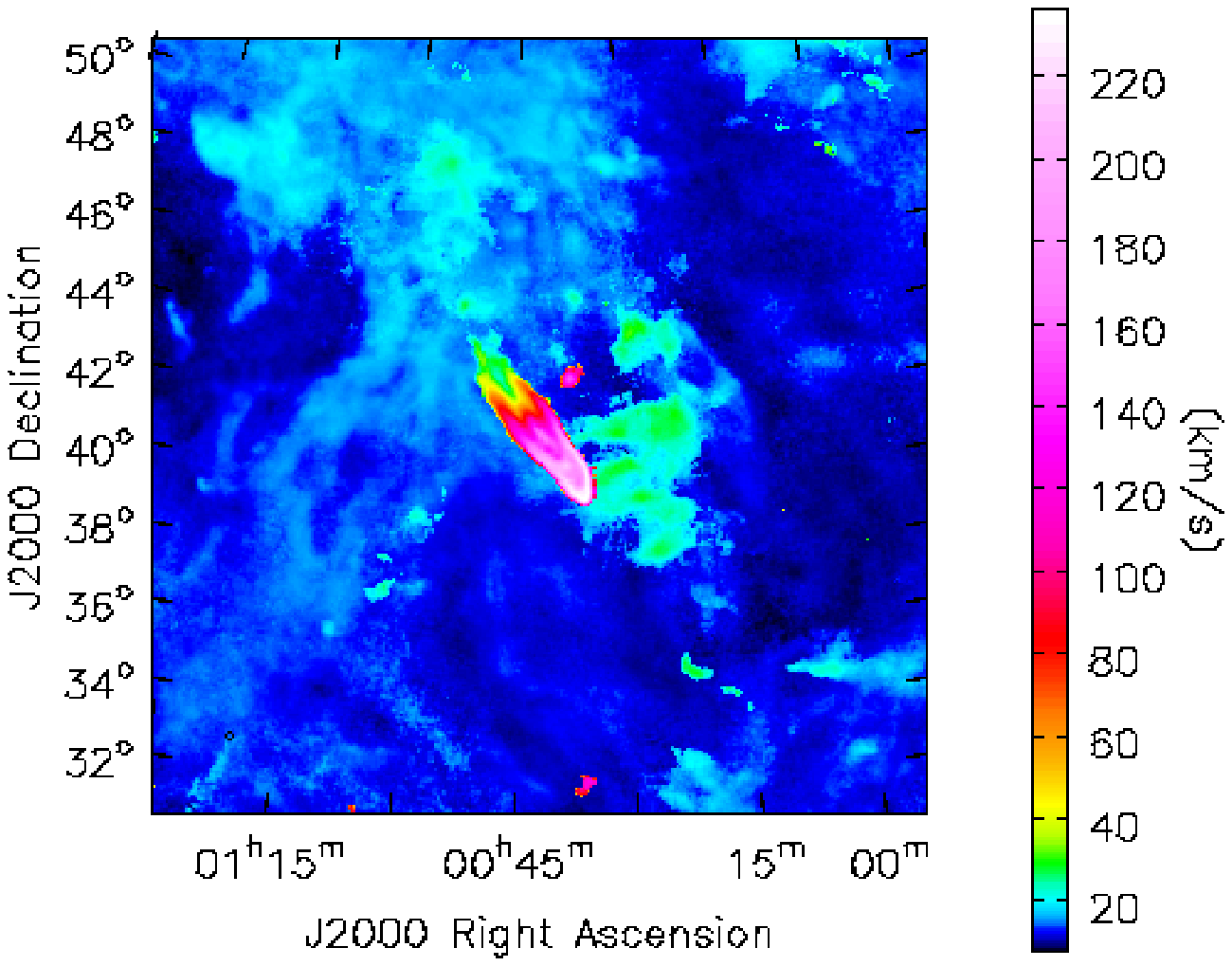}
                }
      \centerline{
      \includegraphics[scale=0.48, angle=0]{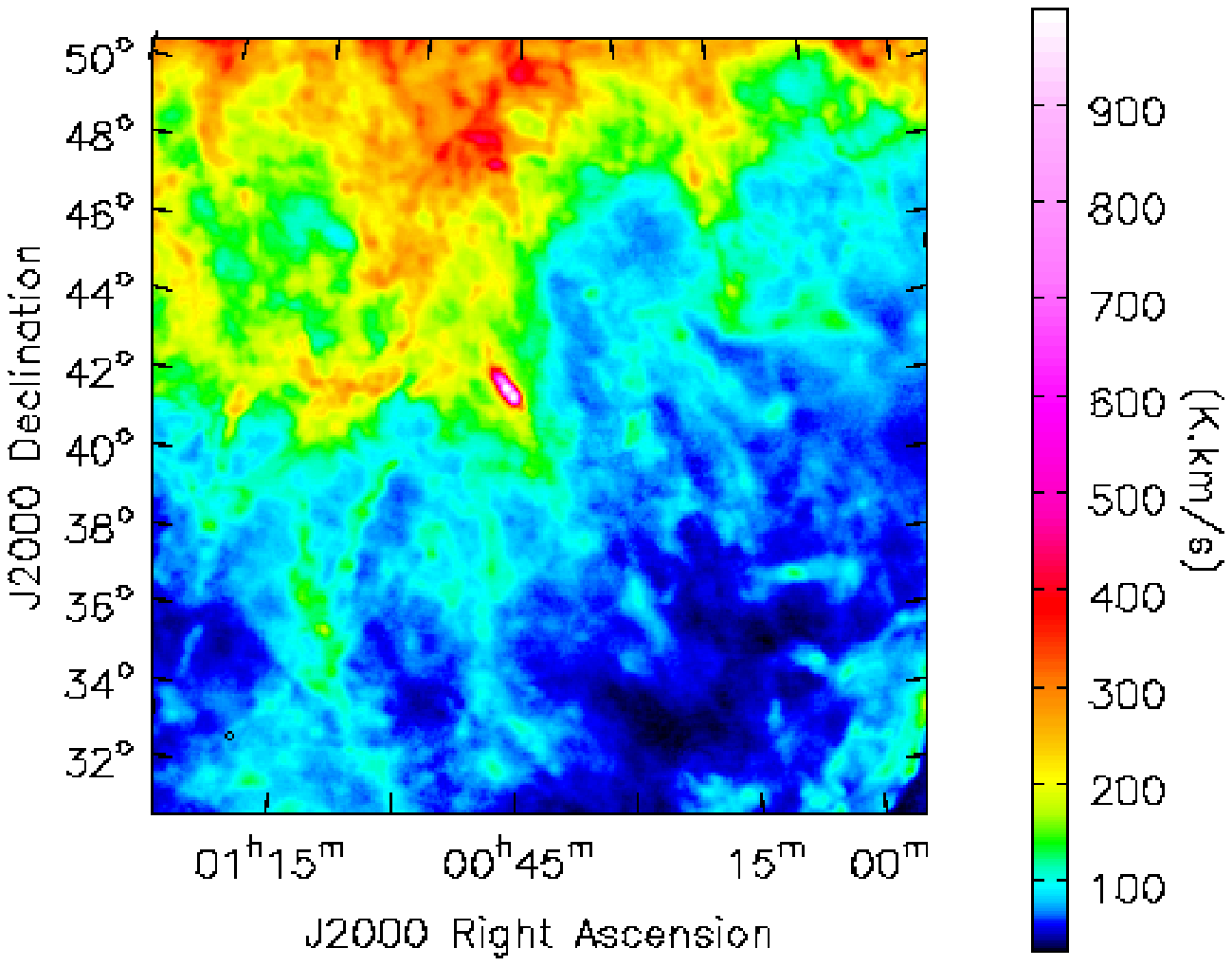}
        \includegraphics[scale=0.48, angle=0]{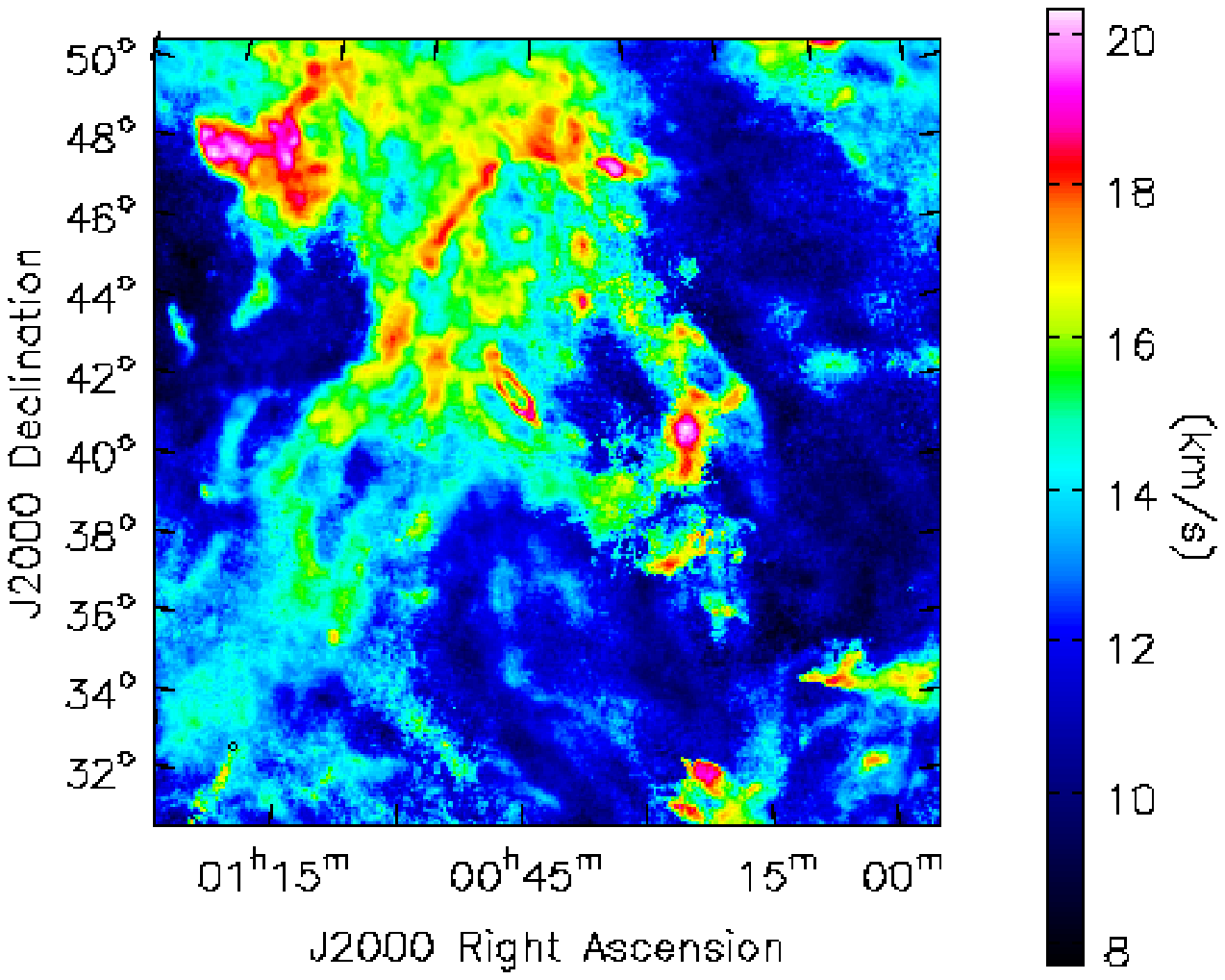}
                }
      
      \caption{Zeroth and second moment maps toward M\,31. {\it Top\/:} The zeroth moment (left) and the second moment map (right) calculated for the radial velocity range $-588\leq\,v_{\rm LSR} {\rm [km\,s^{-1}]}\,\leq -25$ are presented. The disk of M31 is easy to identified in both maps. Prominent in the second moment map are also the lopsided HI structures displayed in Fig.\,\ref{Fig:mom0}. In addition, HI clouds along the bridge toward M33 and isolated HI features like Davies\,cloud \citep{Davies1975} show by their color coding their deviation from the Milky Way galaxy HI emission. {\it Bottom\/:} zeroth moment (left) and the second moment map (right) calculated for the radial velocity range $-75\leq\,v_{\rm LSR} {\rm [km\,s^{-1}]}\,\leq -25$. About 99\% of all HI in the area of interest are comprised in these maps. While the top and bottom zeroth moment maps displayed appear to differ only in the extent of M\,31 HI disk, the second moment maps display totally different structures. We note the very different velocity ranges of both second moment maps and the equal column density range of the zeroth moment maps.
}
    \label{Fig:M31_mom9to450}

    \end{figure*}
\section{Separation of coherent HI Doppler velocity structures by second moment maps} \label{Sect:Separation}
Radio spectroscopic data allow to differentiate not only spatially between different sources of HI emission but also in radial velocity.
In the following we show that the calculation of the so-called ``difference second moment map'' allows one to differentiate individual HI gas features superimposed on a single line of sight.

Moment maps offer information on total intensity,

\begin{equation}
M_0 = \Delta v \sum_{i}^{N} I_i
\label{Eq:mom0}
,\end{equation}

\noindent the velocity field

\begin{equation}
M_1 = \frac{\sum_{i}^{N} I_i \cdot v_i}{M_0}
\label{Eq:mom1}
,\end{equation}
 
\noindent and finally the velocity dispersion field

\begin{equation}
M_2 = \sqrt{\frac{\sum_{i}^{N} I_i (v_i-M_1)^2}{M_0}}
\label{Eq:mom2}
.\end{equation}

In the case of an isolated HI cloud, these maps are used to trace only the information on the physical conditions within this cloud of interest and to give immediate physical meaning.
Equation\,\ref{Eq:mom0} gives the integral across the HI cloud and is a
measure for its column density and eventually for its HI
mass. Equation\,\ref{Eq:mom1} comprises the information on the bulk
motion of the cloud as an entity. Equation\,\ref{Eq:mom2} provides a
measure for the velocity dispersion $\sigma$ of the cloud. In particular, $(v_i-M_1)^2$ is a measure of the velocity deviation along a particular line of sight relative to the average intensity-weighted velocity field $M_1$. In Eq.\ref{Eq:mom2}, $I_i$ weights the velocity irregularities proportional to their HI brightness temperatures. The weighting of $(v_i-M_1)^2$ with $I_i$ is a key to disentangleing accidentally confused coherent HI structures. HI objects showing up with deviating radial velocities $(v_i-M_1)^2$ from the dominant averaged gas ($\propto {M_1}^2\cdot I_i$) become distinguishable in the $M_2$ map as discontinuities.

As an example, in case unrelated radial velocity structures are accidentally superimposed on a single line of sight, the second moment map will show up a $M_2$ discontinuity in comparison to the neighboring unconfused regions. The whole overlapping area will show up with an excess
$M_2$ value that outlines the shape of the overlapping zone. 
The $M_2$ value of two crossing HI filaments, each with a FWHM
of $10\,{\rm km\,s^{-1}}$ and both with the same brightness temperature
$T_{\rm B}$, yields $M_2 = T_{\rm B}\cdot 2\times 4.2\,{\rm km\,s^{-1}}$ at the position of the cross--over  point. Spatially offset from that crossing zone, each individual filament will show up with only half that $M_2$ value. Thus, the second
moment maps mark the location of the cross--over point of the two HI
filaments by doubling the second moment value $M_2$. In case that both HI filaments move with different radial velocities, the $M_2$ value increases proportionally. The discontinuity becomes even more obvious.
But also in the case of a mediocre difference in radial velocities, the $M_2$ value is a sensitive tool for differentiating between both filaments spatially (see Fig.\,\ref{Fig:coldencontrast}).

Because of the weighting of $(v_i - M_1)^2$ with the brightness temperature $I_i$ (see Eq.\,\ref{Eq:mom2}), the column density contrast between the superimposed gas filaments is also important.
Figure\,\ref{Fig:coldencontrast} displays the interrelation between column density contrast and difference in radial velocity of the superposed HI structures. In the case of equal column densities (1:1, meaning filament one and filament two have the same column density), the value of $M_2$ increases rapidly proportional to the radial velocity difference of both filaments. Thus, the cross-over point can be easily identified as a discontinuity in the $M_2$ map even in the case of a small separation of both filaments in radial velocity (in this particular example $\Delta v_{\rm rad} \geq 10\,{\rm km\,s^{-1}}$). In case of a $1:10$ or even $1:100$ column density ratio differentiating between the individual HI filaments demands a much larger separation in radial velocities between both superposed filaments.
This implies that toward high galactic latitudes, with their moderate column densities with $N_{\rm HI} \sim 10^{20}\,{\rm cm^{-2}}$, the second moment map approach can be applied straightforwardly to identify superposed but unrelated HI emission down to $N_{\rm HI} \sim 10^{19}\,{\rm cm^{-2}}$. 

\subsection{The Milky Way galaxy column density distribution toward M31}
According to the considerations above, we have to study the HI column density level of the Milky Way galaxy towards M31 and its velocity structure.
The maximum of the HI column density level toward the area of interest is $N_{\rm HI} = 3.6\cdot 10^{20}\,{\rm cm^{-2}}$ within $|v_{\rm LSR}| \leq 600\,{\rm km\,s^{-1}}$. Eighty percent of this
column density is associated with Milky Way HI gas observed within the
low velocity range $|v_{\rm LSR}| \leq 25\,{\rm km\,s^{-1}}$. Ninety-seven percent of that
diffuse $N_{\rm HI}$ is included by integrating across the intermediate velocity
range $|v_{\rm LSR}| \leq 50\,{\rm km\,s^{-1}}$. Finally 99\% of the total column density across the whole field of interest is contained in the more extended intermediate-velocity  range $|v_{\rm LSR}| \leq 75\,{\rm km\,s^{-1}}$. Thus, the dominant fraction of Milky Way HI gas in the area toward M\,31 is associated with gas at low and intermediate velocities. To explore the superposition of the different HI structures significantly in the $M_2$ map, we exclude the low-velocity regime $|v_{\rm LSR}|\leq 25\,{\rm km\,s^{-1}}$ from further analysis. This choice reduces the column density contrast between the Milky Way and unrelated gas by about 80\%.

Figure\,\ref{Fig:M31_mom9to450} displays the zeroth moment (left panels) and the
second moment maps (right panels).
The moment zero map ($-588\leq\,v_{\rm LSR} {\rm [km\,s^{-1}]}\,\leq -25$, Fig.\,\ref{Fig:M31_mom9to450} top left hand panel) shows M31 but is dominated by Milky Way gas. A Milky Way high column density filament, located close to the northern portion of M31's disk, is obvious. The corresponding $M_2$ map displayed in Fig.\,\ref{Fig:M31_mom9to450} (top right hand panel) reveals a significantly different structure. The disk of M31 is easy to identify as is the southern streams visually emerging from it. Prominent are also the HI clouds
in between M31 and M33 and isolated HI features like the Davies\,cloud \citep{Davies1975}. The $M_2$ values of the second moment map are measures for the FWHM of the individual HI object in addition to their separation in radial velocity. 

The ``dominant average gas'' is determined by calculating the appropriate moments maps $M_0$ and $M_2$ for the radial velocity range $-75\leq\,v_{\rm LSR} {\rm [km\,s^{-1}]}\,\leq -25$. 
Figure\,\ref{Fig:M31_mom9to450} (bottom left) displays the zeroth moment
map of that velocity interval. M31's HI disk emission is only marginally visible next to the
dominating Milky Way high latitude HI structure. Comparing both zeroth
moment maps in Figure\,\ref{Fig:M31_mom9to450} shows almost the same
Galactic high latitude HI column density
structure. Figure\,\ref{Fig:M31_mom9to450} (bottom right) displays the
corresponding second moment map $M_2$. It reveals highly complex and irregular structures. This map is a measure for the line width and the velocity structure of the ``dominant average gas''.

\begin{figure*}
     \centerline{
        \includegraphics[scale=0.5, angle=0]{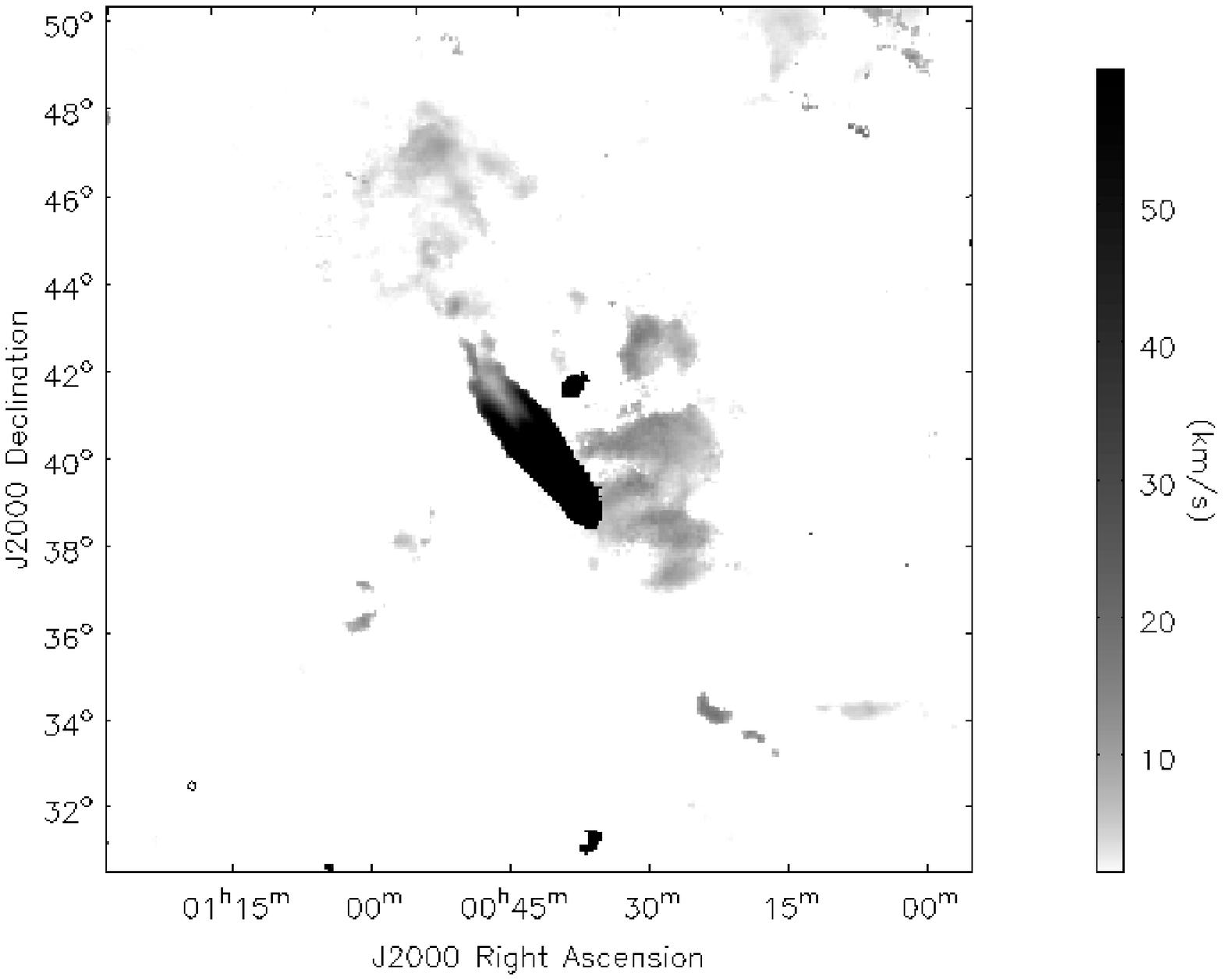}
                }
      \caption{Displayed is the difference second moment map ($\Delta M_2$) calculated from the second moment map of the velocity range $-588\,\leq\,v_{\rm LSR}[{\rm km\,s^{-1}}]\,\leq -25$ and $-75\,\leq\,v_{\rm LSR}[{\rm km\,s^{-1}}]\,\leq -25$ shown in the  righthand panels of Fig.\,\ref{Fig:M31_mom9to450}. We set the lower threshold to $1\,{\rm km\,s^{-1}}$.}
    \label{Fig:M31_mom9to450_400to450}
\end{figure*}

\subsection{The difference second moment map toward M31}
In the next step we subtract the ``dominant average gas'' second moment map (Fig.\,\ref{Fig:M31_mom9to450}, bottom right) from the broad velocity range $M_2$ map (Fig.\,\ref{Fig:M31_mom9to450}, top right). This difference second moment map ($\Delta M_2$) is displayed in Fig.\,\ref{Fig:M31_mom9to450_400to450}.
Most of the area of interest has a $\Delta M_2$ value of exactly zero. These areas indicate the location of Milky Way gas without any confusion with HI gas at more negative radial velocities. 
A high $\Delta M_2$ value is observed toward the top righthand corner, where HI emission associated with HVC complex H is observed (\citet{Simon2006}, $-240 \leq v_{\rm LSR}[{\rm km\,s^{-1}}] \leq -170$). Of particular interest with respect to M\,31, however, is the gas with second moment values up to $10\,{\rm km\,s^{-1}}$. These excess $\Delta M_2$ values are associated with the structure displayed in Fig.\,\ref{Fig:mom0} and apparently
represent the M31\,cloud. The lopsidedness of HI gas around M31 is easily traceable. 
Even faint HI structures covering a broad velocity range are significantly detectable due to their large offset in radial velocity from the ``dominant average gas''.

The M31\,cloud reveals, in particular for the northern HI structure (a), a close association with and a smooth and continuous connection to the M\,31 HI disk
in that map. Individual M31-HVCs that were previously known are apparent \citep[see][Fig.\,2]{Thilker2004}. They all show up with a $\Delta M_2$ deviation value comparable to that of the M31\,cloud features.
This implies that the previously identified M31-HVC gas is a subset of the HI gas highlighted by the difference second moment map $\Delta M_2$. It does not uniquely confirm that all these gaseous structures belong physically to M\,31 but displays the gas not belonging to the Milky Way galaxy. The coherent large scale spatial structure displayed in the $\Delta M_2$ map in combination with Fig.\,\ref{Fig:fourpanels} is a strong argument for the striking hypothesis that the M31\,cloud belongs to M31's HI halo.

Accordingly, the second moment maps give us the chance to distinguish between coherent gaseous structures superposed accidentally and  moving at different radial velocities. The absolute value of that velocity dispersion has no immediate physical meaning but is a measure of the velocity deviation of a specific structure relative to its environment.

\section{Discussion}

\subsection{Absorption lines probing the northern HI structure}
Absorption line spectroscopy contributes additional and independent information on the physical properties of the gas. M31 was the subject of a recent quasar absorption line survey \citep{Rao2013} performed with the Hubble Space Telescope (HST) equipped with the Cosmic Origin Spectrograph (COS). Primary aim of this survey was to probe M31's impact parameter in comparison to the higher redshifted galaxy population. Using the COS data, it is feasible to determine the ionization degree of different heavy elements to evaluate the excitation conditions across a large portion of M31's halo. For our investigation here, the quasar sightline 0043+4234 is of major interest. It probes the northern rim of Andromeda's HI disk. Toward this particular line of sight, the ``deblending'' technique of \citet{Rao2013} localizes the low-ionized Milky Way species at $v_{\rm LSR} = -73\,{\rm km\,s^{-1}}$ and the high-ionized ones at $v_{\rm LSR} = -1\,{\rm km\,s^{-1}}$. The corresponding absorption line complexes attributed to M\,31 are at $v_{\rm LSR} = -233\,{\rm km\,s^{-1}}$ and $v_{\rm LSR} = -191\,{\rm km\,s^{-1}}$, respectively.
Because absorption line spectroscopy is sensitive to column densities orders of magnitudes below the level that we detect in HI emission, the radial velocity limit for the Milky Way gas of $v_{\rm LSR} = -73\,{\rm km\,s^{-1}}$ is really important here. It implies that gas at more negative radial velocities are most likely not physically associated with the Milky Way galaxy.

A dedicated sensitive GBT HI spectrum toward this quasar sight line \citep{Rao2013} is consistent with this finding, because it constrains the Doppler velocity range occupied by M31's HI gas to $-259 \leq v_{\rm LSR} \leq -93\,{\rm km\,s^{-1}}$.
This velocity range obviously includes the HI emission at $v_{\rm LSR} = -115\,{\rm km\,s^{-1}}$ associated with the northern HI structure (a). Gas at more negative velocities than $v_{\rm LSR} = -73\,{\rm km\,s^{-1}}$ appear to belong to M\,31 and its environment and not to the Milky Way galaxy.

Adopting the hypothesis that the whole northern HI structure (a) is physically related to M31, we estimate its linear size to 20\,kpc and its HI mass to $M_{\rm HI} \geq 4.6\cdot 10^7\,{\rm M_\odot}$.  Owing to the limited sensitivity of EBHIS and the applied radial velocity limit for this mass evaluation at $v_{\rm LSR} = -95\,{\rm km\,s^{-1}}$ we could only determine a lower limit. 

\subsection{The southern HI streams (b)}

Inspecting Fig.\,\ref{Fig:mom0} (righthand panel) clearly shows that the velocity structure of the southern HI streams (b) is remarkably different from that of M31's disk.  While this part of M31's HI disk rotates toward the Earth with velocities up to $v_{\rm LSR} = -600\,{\rm km\,s^{-1}}$, the HI streams (b) are detected at much lower velocities around $v_{\rm LSR} = -115\,{\rm km\,s^{-1}}$.
The gap in radial velocity between the M31 disk rotation and the base points of the filaments probably implies that the origin of the southern HI filaments is physically displaced from the approaching side of M31's disk. Also a deceleration of $\Delta v_{\rm LSR} = 20\,{\rm km\,s^{-1}}$ is observed from the base points to the end of the HI streams in the north.

While the HST-COS data \citep{Rao2013} also probe this region of
interest by the quasar 0032+3946 sight line, unique clues cannot be drawn from that data. Blending of the gas absorption features under investigation with the Milky Way galaxy species do not allow the relative contributions to be disentangled. 

\subsection{Is the shape of the M31\,cloud related to the proper motion of M\,31?} 
The proper motion of M31 has been determined by HST
measurements \citep{Sohn2012}. The proper motion direction is oriented approximately from the top right to the bottom lefthand corner in the representation of Fig.\,\ref{Fig:mom0}. Accordingly, the south eastern boundary of the M31 HI disk can be considered as the ``leading edge''. Here, the HI emission is sharply defined along this edge, while the extra-planar M31\,cloud HI emission extends towards the opposite direction. One scenario to form a lopsided HI M31\,cloud is the ram--pressure interaction of M31 with an ambient medium like the warm hot intergalactic medium (WHIM, \citet{Dave2001}). 

According to standard cosmology, the plasma temperature of the WHIM
is assumed to be $T_{\rm WHIM} = 10^6 - 10^7\,{\rm K}$. The HI halo structures identified here have gas pressures around $10\,{\rm K\,cm^{-3}}$, deduced from the HI line width, implying a WHIM volume density of $n_{\rm WHIM} = 10^{-5} - 10^{-6}\,{\rm cm^{-3}}$.

M31's space velocity, relative to the hypothetical WHIM, which is here thought to be at rest in the Local Group's gravitational potential, is $|v| \sim 40\,{\rm km\,s^{-1}}$ \citep{Diaz2014}. 
Even in case of a ($T_{\rm WHIM}, n_{\rm WHIM}$) = ($10^6\,{\rm K}, 10^{-5}\,{\rm cm^{-3}}$) WHIM plasma temperature, volume density combination M31 moves through the WHIM at subsonic speed ($c_{\rm s} \simeq 80\,{\rm km\,s^{-1}}$).
The corresponding WHIM mass densities and M31 proper motion velocity are sufficient
neither to disrupt large amounts of gas from M31 disk nor to trigger enhanced star formation \citep{Steinhauser2012} while in HI morphology an astonishing simililarity to the disrupted HI disks of Virgo cluster galaxies is apparent \citep{Chung2009}.  

\begin{figure}
     \centerline{
        \includegraphics[scale=0.4, angle=0]{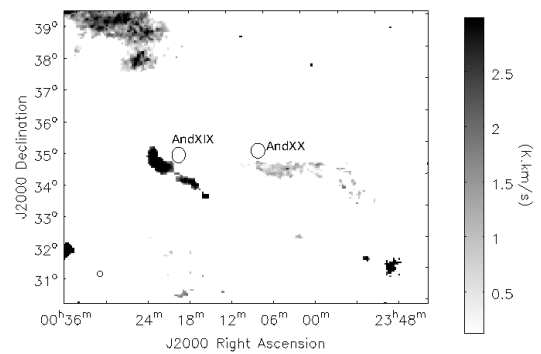}
                        }
      \caption{Integrated ($-478\, \leq\, v_{\rm LSR}\,[{\rm km\,s^{-1}}]\,\leq-117$) brightness temperature map of the area toward Andromeda XIX and XX. All HI emission above 5-$\sigma$ threshold of the EBHIS data with $\Delta v = 12.9\,{\rm km\,s^{-1}}$ is displayed. The circles denote the location of the stellar bodies (30\,arcmin diameter) of both dwarf galaxies according to \citet{Collins2013}.}
    \label{Fig:AndXIX}

\end{figure}
\subsubsection{Displacement between stars and gas of And\,XIX and And\,XX}
While the lopsided Andromeda halo does not allow deriving conclusive evidence for any interaction between the galaxy and an ambient medium, And\,XIX and And\,XX are in this respect of particular interest here.

Figure\,\ref{Fig:AndXIX} shows the EBHIS HI column density distribution toward AndXIX and AndXX. As reported by \citet{GrcevichPutman2009} and \citet{Spekkens2014}, there is a striking non-detection of HI gas toward Milky Way dwarf galaxies but also towardAndromeda's satellite galaxies. Using the EBHIS data we can confirm the non-detection and then consistently lower their limits partly by a factor of two to ten (Kerp et al., in prep). However, these investigations are targeted observations pointing exactly at the stellar bodies of all these dwarf galaxies. The EBHIS column density map shown in Fig.\,\ref{Fig:AndXIX} discloses HI gas close to but not at the stellar bodies of the dwarf galaxies AndXIX and AndXX. The positions of their stellar bodies are marked by the $30\arcmin$ circles in that figure. In the case of AndXIX, the HI gas has almost the same radial velocity
with $v_{\rm LSR} \simeq -115\,{\rm km\,s^{-1}}$  as the stellar component of $v_{\rm LSR} \simeq -112\,{\rm km\,s^{-1}}$ \citep{Collins2013}. The HI gas is displaced from the stellar body and aligned in an extended linear filament.
Also adopting the \citet{Collins2013} distance determination of 820\,kpc for this HI filament yields about $M_{\rm HI}({\rm And\,XIX}) \sim 3.8\cdot 10^7\,{\rm M_\odot}$. 

A similar HI structure is located close to AndXX
(Fig.\,\ref{Fig:AndXIX}). Here, the HI radial velocity ($v_{\rm LSR} = 117\,{\rm km\,s^{-1}}$) is a factor of three lower than that of the stellar body. Only an ensemble of four stars could be identified to be probably associated with AndXX \citep{Collins2013}. Accordingly, improved stellar data of AndXX are important to test this positional correlation with the HI gas.   

 \begin{figure}
     \centerline{
        \includegraphics[scale=0.2, angle=0]{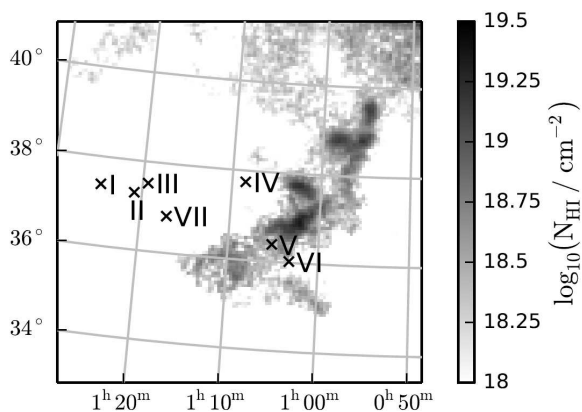}
                        }
          \caption{Integrated column density map ($-600\,{\rm km\,s^{-1}} \leq v_{\rm LSR} \leq -95\,{\rm km\,s^{-1}}$) of the HI emission in between M31 and M33. All HI emission displayed exceeds the
        applied 5-$\sigma$ threshold of the EBHIS data. The crosses mark
        the location of the known HI clouds at systemic velocities \citep{Wolfe2013}. The systemic velocity clouds V and VI closely match the lower velocity clouds in position, while the others are offset by a few degrees. The low-velocity cloud population shows up as a single filament with a pronounced core--halo structure.}
    \label{Fig:bridge_mom0}

\end{figure}

\subsection{The M31/M33 HI bridge cloud population}
Inspecting the whole area shown in Fig.\,\ref{Fig:mom0}, we find a
population of low radial velocity HI clouds between M31 and M33; for a closer look, see also Fig.\,\ref{Fig:bridge_mom0}. These individual clouds are embedded within a common envelope of neutral gas, forming a coherent HI filament. The HI filament is located at and oriented parallel to the northern boundary of the giant stellar stream \citep{Ibata2001, Ibata2007}.

The individual low velocity clouds of that HI filament are located in the same area of the M31/M33 bridge as the known population of systemic velocity HI clouds \citep{Wolfe2013}. Earlier, more sensitive investigations \citep{LockmanFreeShields} toward a neighboring region of interest also report a low-radial velocity HI gas component that has been proposed to be associated with And XV and AndII. However, these HI structures are positionally offset from those reported here, and their brightness temperatures are below our applied 5-$\sigma$ cutoff. The systemic velocity clouds V and VI \citep{Wolfe2013} coincide well in position with the low radial velocity clouds reported here. The systemic population clouds I to IV and VII are located at larger positional offsets.

Recently, the PANDAS field \citep{martin2014} has been investigated again to search for foreground stellar streams out to a distance of 30\,kpc from the Milky Way galaxy. The major aim of their analysis was to distinguish between Milky Way halo features and those belonging to the M31/M33 system of galaxies. None of the already known and the newly identified stellar structures are positionally coincident with the low radial velocity HI filament reported here. Accordingly, the HI filament under investigation does not appear to be associated positionally with any known Milky Way galaxy foreground structure. This suggests there is an  association with Andromeda's intergalactic environment.

In contrast to the HI clouds at systemic velocities, the low-velocity HI filament reveals a pronounced core-halo structure.
The low-velocity clouds cores have HI column densities around $N_{\rm HI} \simeq 1\cdot10^{19}\,{\rm cm^{-2}}$. While the individual HI cores in the filament have comparable brightness temperatures, their linear extents cover a broader range of 3 to $13\,{\rm kpc}$ at Andromeda's distance, corresponding to a HI mass range of $1\cdot 10^5\,{\rm M_\odot}\,\leq\,M\,\leq\,17\cdot10^5\,{\rm M_\odot}$.
We find HI line widths (FWHM) of about $30\,{\rm km\,s^{-1}}$, implying upper limits for kinetic temperatures of $T_{\rm kin} \simeq 2\cdot 10^4\,{\rm K}$.
These values are quantitatively comparable to those of the clouds at
systemic velocities \citep{Wolfe2013}.

\section{Summary and conclusions}
We presented EBHIS first coverage data of the Andromeda's galaxy and its environment. Analyzing channels maps of M\,31 and its environment up to $v_{\rm LSR} = -25\,{\rm km\,s^{-1}}$, we identified a smooth and coherent M\,31 HI halo structure north of the stellar disk. This structure was previously identified by \citet{Blitz1999} as M31\,cloud. Inspecting a difference second moment ($\Delta M_2$) map of the M\,31 and the Milky Way galaxy, we showed that this HI halo structure deviates significantly from the Milky Way HI gas at high galactic latitudes in the velocity regime $-75 \leq v_{\rm LSR}\,[{\rm km\,s^{-1}}] \leq -25,$ which represents the dominant fraction of the HI gas. Quantitatively we used this method to identify all previously known M31-HVC structure with column densities above $N_{\rm HI} = 1\cdot 10^{19}\,{\rm cm^{-2}}$. Accordingly, we proposed a physical association of the M31\,cloud with M31,
first described by \citet{Blitz1999}. The coherent
velocity pattern between Andromeda's tip of the HI disk and the northern
part of the M31\,cloud suggests there is a physical location of this HI
structure in Andromeda's halo. The latest HST absorption line measurements \citep{Rao2013}
are consistent with the association of the M31\,cloud with the Andromeda
galaxy.

Toward the southern portion of Andromeda's HI disk, we observed coherent stream-like HI filaments (b) that originating at the very southeastern edge of the HI disk. These HI streams extend several degrees across the northwestern portion of M31's halo. The brightest portions of these stream-like filaments have been previously reported as individual M31-HVCs by dedicated GBT observations \citep{Thilker2004}.

Assuming that all these HI structures are associated with M31, we
derived a total HI mass of about $M_{\rm halo} = 1\cdot10^8\,{\rm
  M_\odot}$ for M31's extraplanar gas. This is less than \citet{Blitz1999} reported for the M31\,cloud but our mass estimate has to be considered as only a lower limit.

While the well-defined positional coincidence of the sharp boundary of the southern part of the M31\,cloud with the HI rim of Andromeda's southern disk is suggestive of ram pressure interaction, quantitatively the HI streams appear to be too extended to only be formed by that process. The inelastic collision of dwarf galaxies with M31's stellar disk also provide a viable explanation for the existence of the stream-like HI filaments.

It is puzzling that the northern HI structure (a) and the
southern HI streams (b) show up with comparable Doppler velocities. In the case of the different physical origins of both HI structures, this coincidence in radial velocity is unexpected. The same holds true for the low velocity HI filament oriented along the northern rim of the giant stellar stream between M33 and M31.

More sensitive wide-field HI studies of M31 and its environment are needed to construct a consistent view of its HI distribution. In addition to tidal forces and dwarf galaxy impacts into M31's stellar disk, the ram pressure interaction of M31 with its ambient WHIM component might be considered further. High resolution ultraviolet absorption line spectroscopy data are urgently needed to finally clarify the origin of the lopsided giant HI halo of M31.

\begin{acknowledgements}
The authors thank an anonymous referee for very valuable comments eventually leading to the evaluation of the second moment difference map.  We are grateful the Deutsche Forschungsgemeinschaft (DFG) for support of the Effelsberg-Bonn HI Survey project under the grant number KE\,757/7-1 to -3 and KE\,757/9-1. This work is based on observations with the 100-m telescope of the MPIfR (Max-Planck-Institut f\"ur Radioastronomie) at Effelsberg. We like to thank Dr. Michael Bird for careful reading of an earlier version of the manuscript and Prof. Philipp Richter for helpful discussions on the absorption lines.
\end{acknowledgements}

\bibpunct{(}{)}{;}{a}{}{,} 
\bibliographystyle{aa} 
\bibliography{references} 


\end{document}